\documentclass[prb,aps,amsmath,amssymb,reprint]{revtex4-1}
\usepackage{amsmath,amsfonts,epsf,natbib,color,subfloat}
\usepackage{graphicx}
\usepackage{graphics}
\usepackage{subfigure}
\usepackage{epstopdf}
\usepackage{dcolumn}
\usepackage{bm}
\usepackage[mathlines]{lineno}
\usepackage[colorlinks,breaklinks,linkcolor=red,citecolor=blue,urlcolor=blue]{hyperref}

\begin{document}
\title{A linear scaling method to evaluate the ion$-$electron potential of crystalline solids}
\author{Xuecheng Shao}
\altaffiliation{These two authors contributed equally.}
\author{Wenhui Mi}
\altaffiliation{These two authors contributed equally.}
\author{Qiang Xu}
\author {Yanchao Wang}
\email{wyc@calypso.cn}
\author{Yanming Ma} 
\email{mym@calypso.cn}
\affiliation{State Key Laboratory of Superhard Materials, Jilin University, Changchun 130012, P. R. China}
\date{\today}

\begin{abstract}
We propose a simple linear scaling expression in reciprocal space for evaluating the ion--electron potential of crystalline solids. The expression replaces the long-range ion--electron potential with an equivalent localized charge distribution and corresponding boundary conditions on the unit cell. Given that no quadratic scaling structure factor is required---as used in traditional methods---the expression shows inherent linear behavior, and is well suited to simulating large-scale systems within orbital-free density functional theory. The scheme is implemented in the ATLAS software package and benchmarked by using a solid Mg bcc lattice containing tens of thousands of atoms in the unit cell. The test results show that the method can efficiently model large crystals with high computational accuracy.
\end{abstract}


\maketitle


$\emph {Ab initio}$ modeling of materials has become routine in recent years, largely due to the success of density functional theory (DFT).\cite{HK,KS} However, DFT is limited to relatively small systems (about 1000 atoms), and is inappropriate for modeling many atomistic processes---for example fracture or the dynamics of dislocation interactions---where realism is only achieved by considering millions of atoms. Such large-scale simulations are beyond DFT, and require a linear scaling quantum mechanics method. The inherent quasi-linear scaling of orbital-free DFT (OF-DFT) makes it the most promising theory for large-scale simulations.\cite{WangCarter2000,wesolowski2013} In general, all the interaction terms of OF-DFT have linear scaling except for the electrostatic interaction term for periodic systems.\cite{PROFESS,ATLAS} The evaluation of the electrostatic potential therefore is the bottleneck in most OF-DFT programs.\cite{PROFESS,ATLAS}

Generally, the electrostatic potential can be written as the sum of ion--ion, electron--electron, and ion--electron terms. In a periodic system, each of these terms diverges owing to the long-range $1/r$ nature of the Coulomb interaction.\cite{Wigner1997On,Fuchs1935A,Ihm1979Momentum,pask2005real} Divergences that are conditional convergences of extended lattice summations can be eliminated by formulating the summations in terms of neutral densities that are well localized in real/reciprocal space.\cite{Eward} 

The ion--ion term can be transformed to a standard Ewald summation\cite{Eward} under periodic boundary conditions (PBCs), which scales as $\mathcal{O}(M^{2})$, where $M$ is the number of atoms in the system. Owing to the small prefactor of this term, the computational cost is acceptable for OF-DFT calculations of large systems. The electron--electron term can be convoluted in reciprocal space with $\mathcal{O}(N\ln N)$ scaling under PBCs, making its computational cost also acceptable for large-scale simulation. However, the computational cost of the ion--electron potential term of crystalline solids scales as $\mathcal{O}(N\cdot M)$ in reciprocal space owing to the evaluation of the structure factor.\cite{Hung2009Accurate,ATLAS,PROFESS} Here $N$ is the number of gridpoints. Given that the number of gridpoints generally scales linearly with the number of ions, the computational cost of the ion--electron term is effectively $\mathcal{O}(N^2)$ scaling. Note that $N$ is much larger than $M$. Therefore, the ion--electron term dominates the computational time in OF-DFT calculations for large systems.\cite{ATLAS,PROFESS,Hung2009Accurate}

Two methods with much better scaling have been proposed to calculate the ion--electron potential in reciprocal and real space. In reciprocal-space representation, the mathematical trick was employed to significantly reduce the computational cost of calculating the structure factor for large periodic systems.\cite{Hung2009Accurate} The method exhibits linear scaling, and has been successfully applied to systems containing 1 million atoms in the simulated cell.\cite{Hung2009Accurate} In real space representation, a method has been proposed to replace the infinite sum of the long-range Coulomb potential by equivalent localized charge distributions and PBCs. Given the localized charge distributions and the boundary conditions, the summations of all the terms of the electrostatic potential can be evaluated by solving the corresponding Poisson equation.\cite{pask2005real}

Note that the long-range Coulomb potential can be represented as localized ``ion charge'' and the corresponding boundary conditions in the real-space based method.\cite{pask2005real}  Based on this fact, we propose an alternative linear scaling scheme to evaluate the ion--electron potential term of crystalline solids in reciprocal space. Our method can avoid calculation of the structure factor, and thus the method exhibits much better scaling.
In the pseudopotential approximation, the total ion--electron potential $V_{i-e}$ of a crystal can be expressed in real space $V_{i-e}(\bm r)$ or reciprocal space $V_{i-e}(\bm{G})$. Note that $V_{i-e}(\bm r)$ can be simply evaluated by $V_{i-e}(\bm{G})$ with the fast Fourier transform (FFT), which is an $\mathcal{O}(N\ln N)$ operation.\cite{FFT} Therefore, we focus on the expression of ion--electron potential only in reciprocal space. 
For a given periodic system with $n$ atomic species, the total ion--electron potential $V_{\rm i-e}$ can be expressed in reciprocal space as\cite{ATLAS,martin2004electronic}
\begin{equation}
V_{\rm i-e}(\bm G)=\frac{1}{\Omega}\sum_{k=1}^{n}S^{k}(\bm G) V_{\rm loc}^{k}(\bm G),
\label{Vg}
\end{equation}
where $\Omega$ is the volume of the unit cell, $V_{\rm loc}^{k}$ is ionic pseudopotential, and the structure factor of the $k$th atomic species $S^{k}(\bm G)$ is given as 
\begin{equation}
 S^{k}(\bm G)=\sum_{j=1}^{n^{k}}exp(i\bm{G}\cdot {\bm r}_{k,j}),
\end{equation}
where $n^{k}$ and $\bm r_{k,j}$ are the number of atoms and the position of the $j$th atom of $k$th atomic species, respectively. The term $G$ is determined by the primitive vectors of reciprocal space $\bm{b}_{i}$ (i.e., $\bm{G}=n_{1}\bm{b}_{1}+n_{2}\bm{b}_{2}+n_{3}\bm{b}_{3}$, where $n_{i}$ are integers). The evaluation of the structure factor in this expression scales as $\mathcal{O}(N\cdot M)$ instead of $\mathcal{O}(N)$.

\begin{figure}
\includegraphics[width=8.2cm]{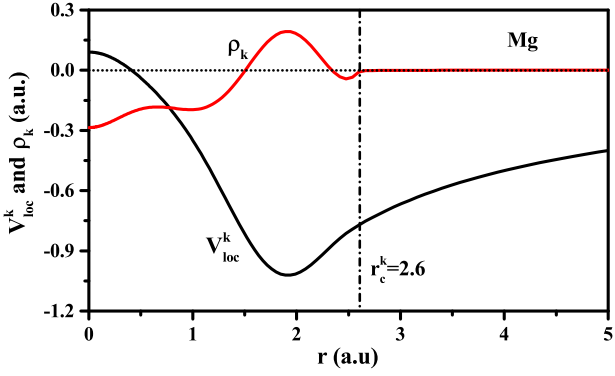}
\caption{Local pseudopotential and the corresponding localized ionic charge density.}
\label{fig:Vloc}
\end{figure}

The local ionic potential of the $k$th atomic species $V_{\rm loc}^{k}(r)$ can be represented by the localized charge density $\rho_{k}(r) $, which can be used to reproduce the equivalent long-range ionic potential. The charge density, $\rho_{k}(r) $, is only localized within the cutoff radius, $r_{c}^{k}$. Fig. \ref{fig:Vloc} shows a typical local ionic pseudopotential and the corresponding localized ionic charge density of Mg, in which the cutoff radius is 2.6 a.u. The spherical symmetry makes the localized charge density 
\begin{equation}
\rho_{k}(r) =\frac{1}{4 \pi}\left(\frac{2}{r}\frac{\partial}{\partial r}+\frac{\partial^{2}}{\partial r^{2}}\right)V_{\rm loc}^{k}(r).
\label{rho} 
\end{equation}

To eliminate the evaluation of the structure factor, the total long-range ion--electron potential $V_{i-e}(\bm G)$ can be obtained by the total ionic charge density, $\rho_{I}(\bm r)$, and the corresponding PBCs. The total ionic charge density can be estimated by summation of all the localized ionic charge density in unit cell.

\begin{equation}
\rho_{I}(\bm r)=\sum_{k=1}^{n}\sum_{j=1}^{n_{k}}\rho_{k}(R_{k,j})
\end{equation}
Here, $R_{k,j}=|\bm r- \bm r_{k,j}|$. 
In principle, $\rho_{I}(\bm r)$ can be used to evaluate the total ion--electron potential in real space $V_{i-e}(\bm{r})$ by solving a Poisson equation with the PBCs.

However, the most convenient way to obtain the ion-electron potential is in reciprocal-space by
\begin{equation}
V_{i-e}(\bm {G})=\left\{
\begin{aligned}
&\frac{4 \pi \rho_{I}(\bm{G}) }{|\bm G|^{2}}\qquad (G\ne 0)\\
&V_{i-e}(G) \qquad(G=0),\\
\end{aligned}
\right.
\label{Vgn}
\end{equation}
where $\rho_{I}(\bm{G})$ can be obtained by the FFT
\begin{equation}
\rho_{I}(\bm G)= FFT(\rho_{I}(\bm r)).
\label{rhoI}
\end{equation}
Just as in the conventional reciprocal method,\cite{ATLAS,martin2004electronic,PROFESS} our method also shows the divergent problem for evaluating ion--electron energy for a charge-neutral periodic system. The problem can be neglected, because the singularity at $G=0$ is canceled exactly by similar divergences in other electrostatic-interaction terms (the ion--ion and electron--electron interactions) in the reciprocal-space representation.\cite{ATLAS,martin2004electronic,PROFESS} The same technique used in the conventional reciprocal method is employed in our scheme. Particularly,
the $V_{i-e}(G=0)$ term in Eq. (\ref{Vg}) can be expressed in reciprocal space as
\begin{align}
V_{i-e}(G=0)&=\frac{1}{\Omega}\sum_{k=1}^{n}S^{k}(G=0)V_{\rm loc}^{k}
(G=0),\nonumber \\
&=\frac{1}{\Omega}\sum_{k=1}^{n}N_{\rm atom}^{k}V_{\rm loc}^{k}(G=0),
\label{Vg0}
\end{align}
where $N_{atom}^{k}$ and $V_{\rm loc}^{k}
(G=0)$ are the number of atoms and the local ionic potential of the $k$th atomic species, respectively. Note that only the difference between the pseudopotenial and the pure Coulomb potential is considered to evaluate the local ionic potential $V_{\rm loc}^{k}$ at $G = 0$. Therefore, the local ionic potential, $V_{\rm loc}^{k}$ at $G = 0$, can be estimated by
\begin{align}
V_{\rm loc}^{k}(G=0)&=4\pi \int_{0}^{\infty}\left(V_{loc}^{k}(r)-(-\frac{Z_{k}}{r})\right)r^{2}dr,
\end{align} 
where $V_{\rm loc}^{k}(r)$ and $Z_{k}$ are the local pseudopotential and the number of the valence electrons of the $k$th atomic species, respectively.  Because
\begin{equation} 
V_{\rm loc}^{k}(r)=-\frac{Z_{k}}{r} \qquad (r > r_{c}^{k}),
\end{equation}  the local ionic potential $V_{\rm loc}^{k}$ at $G = 0$ can be rewritten as
\begin{align}
V_{\rm loc}^{k}(G=0)
&=4\pi \int_{0}^{r_{c}^{k}}\left(V_{loc}^{k}(r)+\frac{Z_{k}}{r}\right)r^{2}dr.
\label{pVg0}
\end{align}
Therefore, $V_{i-e}(\bm G)$ can be determined by Eqs. (\ref{Vgn}) and (\ref{Vg0}).

The detailed processes for evaluating $V_{i-e}(\bm {G})$ are summarized as follows.

i) Evaluate $\rho_{k}(r)$ and $V_{loc}^{k}(G=0)$ via Eqs.(\ref{rho}) and (\ref{pVg0}), and store them in peudopotential files before OF-DFT calculations.

ii) Estimate the total ionic charge density, $\rho_{I}(\bm r)$, by Eq. (\ref{rhoI}) with known $\rho_{k}(r)$ and structural information. 

iii) Calculate $V_{i-e}(\bm G=0)$ and $V_{i-e}(\bm G\ne0)$ by Eq. (\ref{Vg0}) with $V_{\rm loc}^{k}(G=0)$ and Eq. (\ref{Vgn}) with $\rho_{I}(\bm r)$, respectively.

Once $V_{i-e}(\bm G)$ is known, the ion--electron potential in real space, $V_{i-e}(\bm r)$, can be obtained by an inverse FFT
\begin{equation}
V_{i-e}(\bm r)=FFT'(V_{i-e}(\bm G)).
\end{equation} 

To verify the equivalence of the present scheme to the conventional reciprocal-space method, we coded it in Ab initio orbiTaL-free density functionAl theory Software (ATLAS)\cite{ATLAS} and benchmarked it with bulk Mg with a body-centered cubic (bcc) lattice. The TF$+\lambda$vW kinetic energy density functional and the local density approximation exchange--correlation functional parametrized by Perdew and Zunger\cite{ldaca} are used. The local pseudopotential of Mg is constructed by our OEPP scheme\cite{OEPP}, which considers a valence electronic configuration of $3s^{1}3p^{1}$. The core cutoff radius of Mg is set as 2.6 a.u.\cite{OEPP} The ion--ion energy is calculated via Ewald summation.\cite{Eward} 
\begin{figure}
\includegraphics[width=8.2cm]{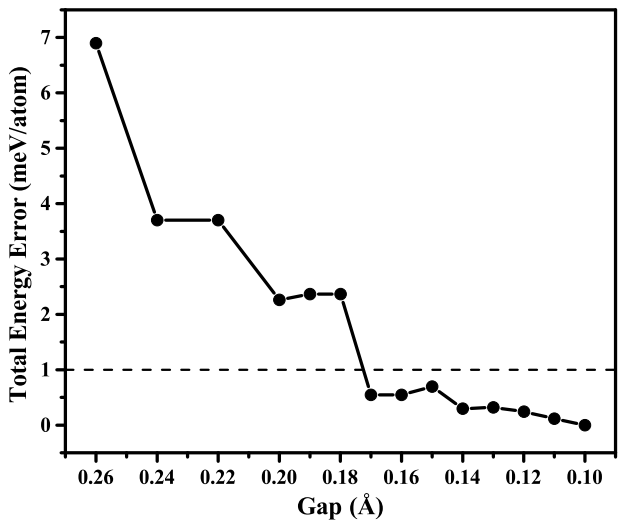}
\caption{Convergence of the total energy of bulk bbc Mg versus the mesh grid gap.}
\label{fig:gap}
\end{figure}

Fig. \ref{fig:gap} shows that the fourth-order finite-difference expansion and a grid spacing of 0.17 \r{A} are sufficient to converge the total energy to well within 1 meV/atom. Therefore, these settings are employed in all the following calculations. Note that a time-saving double-grid technique \cite{Ono1999Timesaving}  is adopted to accurately estimate the total localized ionic charge density for each grid in the unit cell by each ionic localized charge density, which is stored in peudopotential files. The dense grid-spacing is set as $h_{\alpha}^{dens}=h_{\alpha}/2$, and $h_{\alpha}$($\alpha=x,  y$, and $z$) is the coarse grid-spacing. Ninth-order Lagrangian interpolation is used to obtain the charge density of the coarse grid.
\begin{figure}
  \subfigure[]{
  \label{fig:Mg001}
    \includegraphics[width=8.2cm]{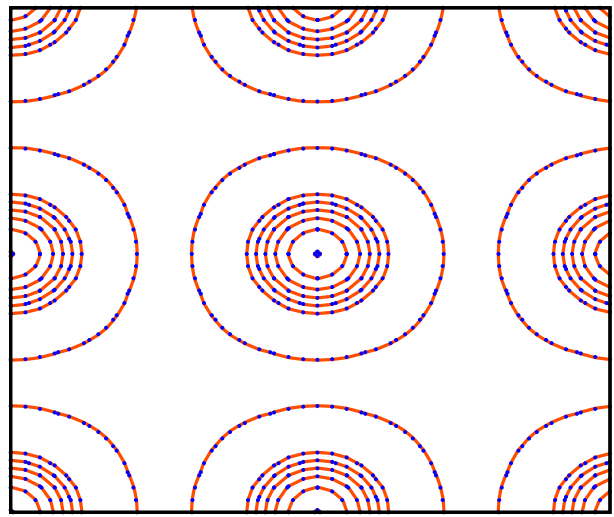}}
  \subfigure[]{
  \label{fig:Mg110}
   \includegraphics[width=8.2cm]{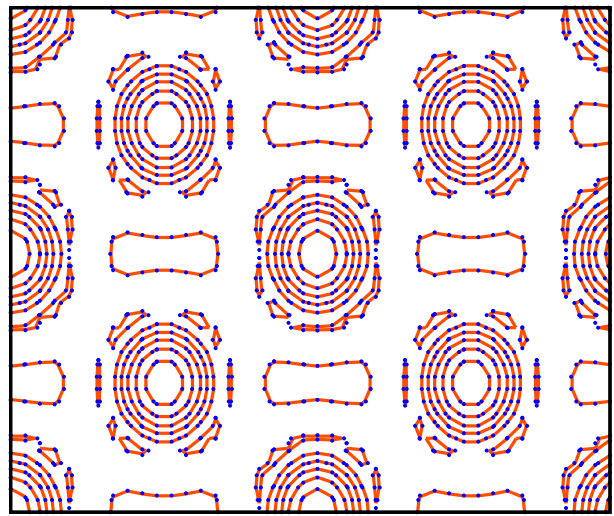}} 
 \caption{Contour plots of ion--electron potential calculated based on the new method (red solid line) and the conventional method (blue dotted line). (a) (001) and (b) (011) planes of Mg with $2\times2\times2$ bcc unit cells.}
\label{fig:density}
\end{figure}

To validate the new scheme, we compare its calculation of an ion--electron potential interaction with that of an exact conventional method. The resulting contour plots (Fig. \ref{fig:density}) show negligible difference in the ion--electron potential of bcc Mg on the (001) and (110) planes calculated by the two methods, demonstrating the accuracy of the new scheme relative to an exact conventional method.
\begin{figure}[!htb]
 \includegraphics[width=8.2cm]{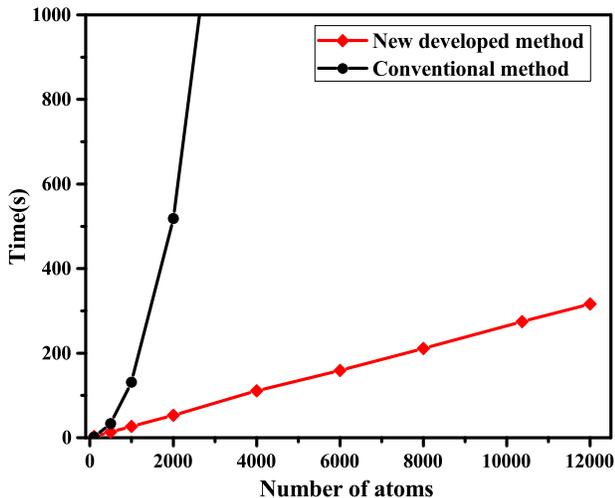}
 \caption{Total time (wall time) to calculate the ion--electron potential term for different numbers of atoms with the conventional method (black line) and the new method (red line).} 
\label{fig:eitime} 
\end{figure}

The performance of our approach is illustrated through its calculation of the ion--electron potential of bulk bcc Mg supercells containing different numbers of atoms (up to 12,000 atoms). The total calculations times of both the conventional method and the new method are shown in Fig. \ref{fig:eitime}. Note the better computational efficiency of the new approach and its approximately linear scaling with system size with a small prefactor due to the advantage of FFT. In particular, the computational time required for an Mg supercell containing 12,000 atoms is decreased substantially from $\sim$18,000 $\emph {s}$ for the conventional method to $\sim$316 $\emph {s}$ for the new scheme.
\begin{figure}
 \includegraphics[width=8.2cm]{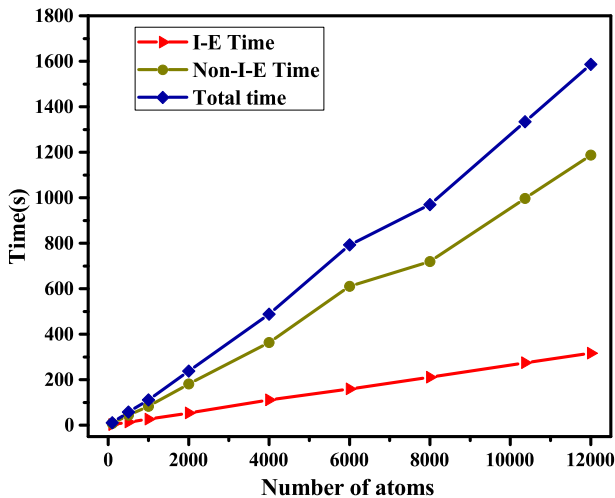}
 \caption{Total time (wall time) using the new method with ATLAS to calculate the total energy during electron-density optimization for systems of 100 to 12,000 atoms in a simulated bcc Mg cell. The total time (blue line) is shown as the sum of the times for the ion--electron potential term (red line) and for all other potential and energy terms (green line).}
\label{fig:time}
\end{figure}

Within this scheme, the computational efficiency of ATLAS\cite{ATLAS} is tested further on supercell Mg with a single processor. The total time and its contributions from the time to calculate the ion--electron term and all other terms throughout the electron density optimizations are presented in Fig. \ref{fig:time} for systems containing 100 to 12,000 atoms. All terms show approximately linear scaling as the number of atoms increases owing to the linear scaling method used to calculate the ion--electron potential term. The proportion of time spent calculating the ion--electron term is trivial, and does not dominate the total computational time within the new scheme. In this regard, our new scheme can greatly improve the computational efficiency of ATLAS the software, and could be applied to large-scale OF-DFT simulations.

In summary, an alternative simple expression for calculating the ion--electron potential of crystalline solids is proposed. Because the expression does not require evaluation of the structure factor for periodic systems, our approach shows linear scaling and can effectively overcome the limitation of high computational cost of conventional approaches. Therefore, it is well suited to simulating large-scale systems within OF-DFT. The method is implemented in ATLAS software and benchmarked using bcc Mg containing large numbers of atoms per unit cell (up to 12,000 atoms). The results show that our method can achieve high computational accuracy and efficiency.


Y.M., Y.W., X.S., and W.M. acknowledge the funding support from the National Natural Science Foundation of China under Grant Nos. 11274136, 11534003, and 11404128 and from the 2012 Changjiang Scholar of the Ministry of Education and the China Postdoctoral Science Foundation (No. 2015T80294 and No. 2014M551181). 

\nocite{}
\bibliography{IE_potential}
\end{document}